\documentstyle[aas2pp4,psbox]{article}

\lefthead{Shirey, Levine, \& Bradt}
\righthead{Absorption Dips in Cir X-1}

\newcommand{\RXTE}{{\it RXTE\,}}
\newcommand{\EXOSAT}{{\it EXOSAT\,}}
\newcommand{\ASCA}{{\it ASCA\,}}
\newcommand{\ROSAT}{{\it ROSAT\,}}
\newcommand{\Ginga}{{\it Ginga\,}}
\newcommand{\intens}[2]{I\,[#1--#2~keV]}

\newcommand{\hardness}[4]{\intens{#3}{#4}~/ \intens{#1}{#2}}
\newcommand{\msun}{${\rm M}_\odot$}
\newcommand{\scinot}[2]{${#1}{\times}10^{{#2}}$}
\newcommand{\mscinot}[2]{{{#1}{\times}10^{{#2}}}}
\newcommand{\tento}[1]{$10^{{#1}}$}

\begin{document}

\title{Scattering and Iron Fluorescence Revealed During Absorption Dips in Circinus~X-1}
\author{
Robert E. Shirey\altaffilmark{1,2}, 
Alan M. Levine,
Hale V. Bradt\altaffilmark{1}}

\affil{
Center for Space Research, Massachusetts Institute of Technology, 
Cambridge, MA 02139 \\
{\bf Accepted for publication in The Astrophysical Journal}\\
}

\altaffiltext{1}{Department of Physics, Massachusetts Institute of
Technology, Cambridge, MA 02139}

\altaffiltext{2}{
Present address: Department of Physics, University of California, Santa Barbara,
\\ Santa Barbara, CA, 93106; shirey@orion.physics.ucsb.edu}

\begin{abstract}

{\em Rossi X-ray Timing Explorer} (\RXTE) All-Sky Monitor light curves
of Circinus~X-1 show that intensity dips below the $\sim$1~Crab
baseline occur near phase zero of the 16.55-d cycle of the source.
\RXTE\ Proportional Counter Array observations carried out between
1996 September~20--22 provided 60\% observing efficiency for 48~hours
around phase zero. These observations showed significant dipping
activity during much of those two days.  The dramatic spectral
evolution associated with the dips is well-fit by variable and at
times heavy absorption ($N_H > $ \tento{24}~cm$^{-2}$) of a bright
component, plus an underlying faint component which is not attenuated
by the variable column and whose flux is about 10\% of that of the
unabsorbed bright component.  A prominent iron emission line at
6.4--6.6 keV is evident during the dips.  The absolute line flux
outside the dips is similar to that during the dips, with equivalent
width increasing from as low as 44~eV outside dips to more than 400~eV
during dips, indicating that the line is associated with the faint
component.  These results are consistent with a model in which the
bright component is radiation received directly from a compact source
while the faint component may be attributed to scattered radiation.
Our results are also generally consistent with those of Brandt et~al.,
who found that a partial-covering model could explain spectra obtained
in \ASCA\ observations of a low-to-high transition in Cir~X-1.  The
relative brightness of the two components in our model requires a
column density of $\sim$\scinot{2}{23}~cm$^{-2}$ if the faint
component is due to Thomson scattering in material that mostly
surrounds the source.  We find that illumination of such a scattering
cloud by the observed direct component would produce an iron K$\alpha$
fluorescence flux that is in rough agreement with the flux of the
observed emission line.  We also conclude that if the scattering
medium is not highly ionized, our line of sight to the compact source
does not pass through such a scattering cloud.  Finally, we discuss
simple pictures of the absorbers responsible for the dips themselves.

\end{abstract}


\section{Introduction}

Circinus~X-1 exhibits complex X-ray variability during its 16.55-d
intensity cycle (e.g., \cite{kaluzienski76}; Dower, Bradt, \& Morgan
1982\nocite{dower82}; \cite{tennant87}, 1988\nocite{tennant88};
\cite{stewart91}, Shirey et~al.\ 1996, 1998, hereafter Papers
I\nocite{shirey96} \& II\nocite{shirey98:feb97}; Shirey, Bradt, \&
Levine 1999, hereafter \cite{shirey99:fullZ}). Enhanced mass transfer
near periastron of a highly eccentric binary orbit may be responsible
for both flaring and dipping behavior (\cite{murdin80};
\cite{oosterbroek95}; \cite{brandt96}). Three type~I X-ray bursts seen
from Cir~X-1 during an \EXOSAT\ observation demonstrate that the
system contains a neutron star with a weak magnetic field (Tennant,
Fabian, \& Shafer 1986\nocite{tennant86}).

{\em Rossi X-ray Timing Explorer} (\RXTE) All-Sky Monitor (ASM;
2--12~keV) light curves (see Papers I\nocite{shirey96} \&
II\nocite{shirey98:feb97} and \cite{shirey98:phd}) show that Cir~X-1
has maintained a bright baseline level of $\sim$1~Crab (1060~$\mu$Jy
at 5.2~keV) since early 1996. Dramatic flares generally begin during
the day following phase zero, as given by the radio ephemeris
(\cite{stewart91}), of each 16.55-d cycle of the source. The flaring
reaches as high as 3.7~Crab and typically lasts 2--5 days, but
sometimes persists for most of the cycle. This behavior is similar to
that observed in folded \Ginga\ ASM data (\cite{tsunemi89}).

The \RXTE\ ASM light curves of Cir~X-1 also show intensity points
below the 1~Crab baseline near phase zero of about half of the cycles
observed (see Fig.~3-1 of \cite{shirey98:phd}).  Because ASM coverage
is incomplete (about ten to twelve 90-s exposures per day), it is
possible that dips, lasting seconds to hours in duration, might
actually occur in all cycles.  During such dips, the X-ray intensity
reaches as low as $\sim$80~mCrab.  The dips most often occur during
the half day before and after phase zero, but sometimes occur up to
two days after phase zero.  Occasionally, dips also occur at later
phases in the ASM light curve, e.g., near day~4 of the cycle.

X-ray dips are common in the light curves of X-ray binaries.  In some
cases, dips may be due to a true decrease in the X-ray emission, e.g.,
due to an accretion disk instability in which the inner part of the
disk is ejected (\cite{greiner96}; \cite{belloni97}).  However, in
most cases dips are attributed to absorption by matter passing through
the line of sight to the X-ray emitting region (see \cite{white95} for
a recent review).  Deep, erratic dips indicate a compact emission
region and structure in the absorber, probably near the outer edge of
the accretion disk.  Shallower, smooth dips occur in systems where the
compact source is always obscured by the disk (due to a high
inclination) so that only partial covering of an extended corona can
occur.  In some systems, X-ray eclipses by the companion star occur
and are total or partial depending on whether the non-eclipse emission
is dominated by a compact source or an extended region such a corona.

An \ASCA\ observation of Cir~X-1 at phase zero showed a low-to-high
intensity transition (\cite{brandt96}), apparently the egress from a
dip similar to those seen with the ASM\@. Brandt et~al.\ used a
partial-covering model to fit the spectra before and after the
transition. A strong iron K~edge in the low-state spectrum indicated
that the obscuring matter had a very high column density, i.e., 
$N_H > 10^{24}$~cm$^{-2}$. Those authors suggested that scattering by
electrons in circumstellar material might be the source of the faint
(dip) spectrum.

In Papers I\nocite{shirey96}, II\nocite{shirey98:feb97}, \&
III\nocite{shirey99:fullZ} we presented \RXTE\ results on the spectral
and timing evolution of Cir~X-1 during quiescent and active phases,
exclusive of the behavior during dips.  Those results demonstrate that
Cir~X-1 at least sometimes exhibits Z-source behavior, including
horizontal branch oscillations (HBOs) at unusually low frequencies
(1.3--35~Hz), normal branch oscillations (NBOs) at $\sim$4~Hz, and
only very low frequency noise (VLFN) on the flaring branch. In this
fourth paper we focus on spectral evolution during dips observed with
the \RXTE\ Proportional Counter Array (PCA).

\section{Observations}

Observations of Cir~X-1 with the PCA were made during 1996
September~20--22, covering 48-hours around phase zero with $\sim$60\%
observing efficiency (see Fig.~\ref{fig:48hr}). These
observations showed significant dipping for about half of the observed
time and provided the opportunity to study dip behavior in detail.

The observations began about a half day before phase zero.  The
intensity of Cir~X-1 was very low (an extended dip) during most of the
first half day of the observations, and then returned to the much
higher ``quiescent'' level shortly after phase zero.  Shorter
intermittent dips occurred during the $\sim$0.3 days following phase
zero.  The strongest dips all reached similar minimum levels of about
200--350 counts~s$^{-1}$~PCU$^{-1}$ (2--18~keV) after background
($\sim$10 c~s$^{-1}$~PCU$^{-1}$) was subtracted, where 1~Crab
$\approx$ 2600~c~s$^{-1}$~PCU$^{-1}$\@.  A second episode of
significant extended dipping occurred from about 0.5 to 0.9~days after
phase zero, followed by ``active-state'' behavior for the remaining
half day of the observation.  The active state is characterized by (1)
a gradual increase of the average intensity in the lowest-energy PCA
band (which dominates the total PCA count rate) accompanied by a
significant drop in the high-energy bands, and (2) individual flares
on shorter timescales (less than a few hours) which show increased
intensity in all PCA energy bands.  The former behavior results in
gradual shifting of spectral tracks in color-color and
hardness-intensity diagrams (CDs/HIDs) and is related to the 16.55-d
intensity cycle of Cir~X-1.  The latter behavior corresponds to motion
along branches of the ``Z'' track in CDs and HIDs. For example, based
on tracks in hardness-intensity diagrams and on timing properties, we
find that the burst-like events near day 348.0 in
Figure~\ref{fig:48hr} correspond to motion up the flaring branch. The
final two segments of Figure~\ref{fig:48hr} (after day 348.1) show
motion from the NB/FB apex up the normal branch to higher total
intensity, and back down. The spectral and timing evolution during the
active state of Cir~X-1 was discussed in detail in Papers
II\nocite{shirey98:feb97}~\&~III\nocite{shirey99:fullZ}.

\section{Evolution of Hardness Ratios During Dips}

Three dips, each showing a complete transition from the baseline
intensity level down to a deep minimum, have been selected for further
study and are labeled Dips 1--3 in Figure~\ref{fig:48hr}.  An expanded
2400-s segment of Figure~\ref{fig:48hr} is shown for each of these
dips in Figures \ref{fig:dip1lc}, \ref{fig:dip2lc},
and~\ref{fig:dip3lc}.  The light curves in multiple PCA energy
channels show that the transitions into and out of dips occur more
rapidly at low energy that in higher energy bands; this is most
pronounced for Dip~3 in Figure~\ref{fig:dip3lc}.  The hardness ratios
thus initially increase (harden) as the denominator falls more quickly
than the numerator. Low-energy count rates are also the first to reach
a minimum level well above background. This causes the hardness ratios
to decrease (soften) since the denominator approaches a constant while
the numerator continues to decrease.

The intensity and hardness ratios from Dip~3 have been used to produce
the color-color and hardness-intensity diagrams (CD \& HID) shown in
Figure~\ref{fig:cc_hid_85911900_2ks}. The spectral evolution during
the dip produces tracks with dramatic bends in both the CD and the
HID.  A similar CD and HID (Fig.~\ref{fig:cc_hid_85823000_17ks}) has
also been produced for the first three segments of
Figure~\ref{fig:48hr} (day 346.31--346.52), during which the intensity
was mostly in an extended low/dip state but briefly reached higher
intensity.

In the HIDs of Figures~\ref{fig:cc_hid_85911900_2ks}
and~\ref{fig:cc_hid_85823000_17ks}, dips produce motion to the left
(toward lower intensity) and initially upward (harder), but turn
downward (softer) as the count rate in the denominator approaches its
bottom level.  In the CDs where, by definition, two hardness ratios
(hard and soft color) are employed, motion is initially to the right
and upward (harder in both colors but mostly in the lower channels).
When the lowest channel reaches the bottom level, the track turns to
the left but continues upward. When intermediate channels stop
decreasing and only the highest channels still drop, the track finally
turns downward. The intensity of the second and third lowest energy
bands rapidly drop by a factor of two at time 1900~s in
Figure~\ref{fig:dip3lc}, just before settling to their lowest levels;
this resulted in a gap in the the middle CD branch in
Figure~\ref{fig:cc_hid_85911900_2ks}.

Evolution of the position in CDs and HIDs apart from dips is
interpreted in Papers~II\nocite{shirey98:feb97} \&
III\nocite{shirey99:fullZ} as motion around a ``Z'' track as well as
shifts of the entire ``Z'' pattern. The exact position of a dip track
in the diagrams depends on its starting point, i.e., the baseline
intensity and spectrum.  Although starting from different positions,
all dips below the 1-Crab baseline in PCA light curves of Cir~X-1
produce the same general shape in CDs and HIDs. However, the tracks
may not be complete for weak dips.

\section{Evolution of Energy Spectra During Dips}

\subsection{Spectral Model}

Cold (neutral) matter most effectively absorbs photons at lower
energies, so gradually increasing obscuration by cold matter produces
dips that are more gradual at high energy than at low energy, as we
observe in Cir~X-1. However the fact that strong dips in Cir~X-1 show
a minimum intensity well above background suggests the presence of a
faint component unaffected by the dips. Thus, in modeling spectra of
dips in Cir~X-1, at least two components are necessary: (1) a bright
component modified by variable and at times very heavy absorption and
(2) a faint component seen through a constant and relatively low
column density. 
From a study of the energy spectrum during non-dip observations
(\cite{shirey99:fullZ}), we found that the 2.5--25~keV spectrum of
Cir~X-1 was generally fit well with a model consisting of a
multi-temperature ``disk blackbody'' plus a higher-temperature
($\sim$2~keV) blackbody.  We find that a model based on these
components also adequately fits the current spectra, so a disk
blackbody plus blackbody is adopted for the bright component during
dips.
For simplicity, we assume that the unabsorbed spectral shape of the
faint component is similar to that of the bright component, as might
be the case if it is due to scattering, e.g., by a corona or stellar
wind, of radiation coming directly from a compact central source back
into the line of sight from other initial emission directions. 
Thus for the faint component we use the same model and parameters as
the bright component, but normalized by an energy-independent
multiplicative factor $f$ which is less than unity.  We discuss the
validity of this assumption below.

The absorption column density for the faint component is assumed to
remain constant throughout a dip, but the column density for the
bright component is allowed to vary to produce the dips.  The light
curves during strong dips show a significant reduction in intensity
even at high energy (13--18~keV), indicating very high column
densities ($ N_H > 10^{24}$ cm$^{-2}$). Photoelectric absorption is
the dominant process responsible for dips at low energy, but the
photoelectric cross-section decreases with energy. Thus, above about
10~keV, Thomson scattering (with an approximately energy-independent
cross section) becomes dominant. Both photoelectric and
Thomson-scattering factors are included in the attenuation calculation
for the bright component.  For simplicity, the effective hydrogen
column density and the Thomson-scattering (electron) column density
are taken to be equal; the actual value of the latter will be slightly
higher (by $\sim$20\%) due to atoms (mainly helium) with more than one
electron.  The complete model used can then be expressed as:
\begin{eqnarray} 
\label{eq:dip_model}
F = 
\left[\exp(-\sigma_{ph}{N_H}^{(1)})\exp(-\sigma_{Th}{N_H}^{(1)})\right] M + 
\nonumber \\
	\left[\exp(-\sigma_{ph}{N_H}^{(2)}) f \right] M,
\end{eqnarray}
where $F$ is the observed flux, $\sigma_{ph}$ and $\sigma_{Th}$ are
the photoelectric and Thomson cross-sections, ${N_H}^{(1)}$ and
${N_H}^{(2)}$ are the effective hydrogen column densities of the
bright and faint components respectively, $f$ is the ratio of the
unabsorbed flux of the faint component to the unabsorbed flux of the
bright component, and $M$ is the disk blackbody plus blackbody model.
For all model components we assume solar abundances (\cite{anders82}),
and we used the photoelectric cross-sections given by Morrison \&
McCammon (1983\nocite{morrison83}).  

\subsection{Energy Spectra Inside and Outside Dips}

We make use of joint fits of spectra from inside and outside a given
dip to simultaneously constrain the parameters of the various spectral
components.  The total $\chi^2$ value obtained when the model is fit
to multiple spectra is minimized by varying the model parameters, but
allowing only the absorption column density of the bright component to
vary among spectra.  Since the disk blackbody and blackbody parameters
typically evolve on timescales of hours (see \cite{shirey99:fullZ}),
it is important in this analysis to use a spectrum from immediately
before (or after) the dip.

The spectral fits were carried out on background-subtracted PCA
pulse-height spectra.
A 1\% systematic error estimate was added in quadrature to the
estimated statistical error (1~$\sigma$) for each channel of the
spectra to account for calibration uncertainties. Fits were carried
out only on data from PCUs 0, 1, and 4 and for energy channels
corresponding to 2.5--25~keV, since the response model (version 2.2.1)
for the other two PCA detectors and for outside that energy range has
been determined to be less accurate (\cite{remillard98}). Fit
parameters reported are the average values for PCUs 0, 1, and
4. Errors are conservatively estimated as the entire range encompassed
by the 90\% confidence intervals from each of the detectors.


Time segments immediately prior to dips 1 and 2 (``A'') and during the
lowest portions of dips 1 and 2 (``B'') were selected for spectral
analysis (dashed lines in Figs.~\ref{fig:dip1lc} and
\ref{fig:dip2lc}). The two segments from Dip~1 were each 96~s
in duration, and those from Dip~2 were 304~s in duration. The
pulse-height spectra from these two time segments are shown in Figure
\ref{fig:dips1and2_nogauss} along with best-fit models.  The
parameters for the bright and faint component of each best-fit model
are listed in the first two lines of Table~\ref{tab:continua}.
The distance to Cir~X-1 has been estimated to be about 6--10~kpc
(\cite{stewart91}; \cite{goss77}), so we chose a value of 8~kpc in
converting blackbody and disk blackbody normalizations to radii.  The
column density (${N_H}^{(2)}$) of the faint component is given in the
table. The column density (${N_H}^{(1)}$) of the bright component
differs for spectra A and~B of each fit and is thus given separately
in Table~\ref{tab:absorpAB}.

The disk blackbody and blackbody parameters for both dips are similar
to those obtained for non-dip spectra in \cite{shirey99:fullZ}
(temperatures of 1.5~keV and 2.2~keV respectively). The unabsorbed
flux of the faint component is $\sim$10\% that of the bright
component. The column density of the faint component is very low, and
consistent with zero. This low value is inconsistent with estimates of
the interstellar column density ($N_H = $ 1.8--2.4$\times
10^{22}$~cm$^{-2}$) measured with \ASCA\ and \ROSAT\ (\cite{brandt96};
\cite{predehl95}).  Apparently, the spectral shape of the faint
component is slightly softer than that of the bright component, giving
the appearance of reduced interstellar absorption.  In fact, the
\ASCA\ data from a dip egress also showed that the low-state spectrum
was softer than the high-state spectrum, and Brandt et~al.\
(1996\nocite{brandt96}) hypothesized that this may be due to X-rays
scattered into the line of sight by interstellar dust. This scattering
by dust is distinct from the scattering by circumstellar material that
might be responsible for the low-state spectrum itself.  The use of
different spectral models for the bright and faint components or the
addition of another continuum component would not be productive due
to the large number of parameters involved, so we allow the column
density of the faint component to compensate for small differences in
spectral shape.

The column density of the bright component outside dips (spectra 1A
and 2A; see Table~\ref{tab:absorpAB}) is about
\scinot{3}{22}~cm$^{-2}$, just slightly above the interstellar
value. The column density during the low segments of dips (spectra 1B
and 2B) is extremely high, i.e., \scinot{176}{22}~cm$^{-2}$ for spectrum~1B
and \scinot{306}{22}~cm$^{-2}$ for spectrum~2B\@. This high column
density value is required to produce the observed reduction in flux at
20~keV and is sufficient to render the absorbed component totally
negligible at lower energies. Thus, the flux in spectra 1B and~2B
below 5~keV may be attributed entirely to the faint component.

\subsection{Iron Emission Line}

Pulse-height spectra from the bottom of dips (spectra 1B and~2B in
Fig.~\ref{fig:dips1and2_nogauss}) show a prominent peak near
6.5~keV\@, suggesting iron K$\alpha$ emission.  The residuals for
fitted spectra from both inside and outside the dips are similar, and
in particular show peaks near 6.5~keV with similar absolute flux
levels.  This suggests that the line feature is associated with the
faint component. If the faint component is visible primarily because
of scattering, iron fluorescence would be expected to occur in the
scattering medium (see discussion section below).

The fits were repeated including a Gaussian emission line component at
6.4--6.6~keV\@.  Allowing the line flux to differ between fits of
spectra A and B of each dip confirms that the line flux is the same
to within a factor of two or less, despite an order of magnitude
difference in continuum flux. Thus we jointly fit the spectra of each
dip using the same line parameters for both spectra
(Fig.~\ref{fig:dips1and2_gauss}).  The addition of the line does
reduce the magnitude of the residuals near 6.5~keV, but does not
improve the fit below $\sim$5.5~keV\@.  The results of the joint fits,
including the Gaussian line, are listed in the second group in
Tables~\ref{tab:continua} and \ref{tab:absorpAB}. They are generally
consistent with the model parameters from the previous fits within the
90\% confidence limits. Table~\ref{tab:continua} also shows that a
significant improvement in the reduced $\chi^2$ value was achieved in
both cases. We will continue to include a Gaussian line in all
subsequent fits and list its fitted parameters in
Table~\ref{tab:gauss}. Although the best-fitting centroid energy of
the line is close to 6.6~keV, the relatively coarse energy resolution
of the PCA ($\sim$1~keV FWHM at 6~keV) is insufficient to confirm
whether the line is from neutral or partially ionized iron or whether
it is actually composed of multiple unresolved narrow lines.

\subsection{Spectral Evolution During Dip Transitions}

Through its large collecting area, the PCA provides good-quality
spectra of bright sources such as Cir~X-1 every 16~s. This enables us
to study the detailed evolution of the spectrum during the transitions
between high and low flux states outside and inside dips.  We selected
four to six 16-s segments from the ingress of dips 1, 2 and 3
(indicated as circles on Figures \ref{fig:dip1lc}, \ref{fig:dip2lc},
and~\ref{fig:dip3lc}) and jointly fit the 4--6 spectra of each dip
with the model developed above.  The 16-s pulse-height spectra from
each of these dips are shown in Figures \ref{fig:dip1_four16s},
\ref{fig:dip2_four16s}, and~\ref{fig:dip3_six16s}, together with the
model curves and residuals.

The low-energy intensity in each case initially decreases but then
reaches a fixed level while the intensity at higher energy continues
to decrease.  A ``step'' near 7.1~keV is evident in spectra (most
prominently in spectra 1c, 2c, 3d) for which significant, but not
total, absorption of the bright component occurs at that energy. This
step cannot be accounted for solely by the emission line but is
naturally fit by iron K-edge absorption intrinsic to neutral or
nearly-neutral absorbing material with a column density (assuming
solar abundances) of $N_H >$ few$\times10^{23}$ (intermediate and
lower spectra, Table~\ref{tab:absorp16s}).  As noted above, the
intensity at the bottom of the dip is significantly reduced relative
to outside the dip even at 20~keV\@.

The fit parameters for the bright and faint components in each joint
fit of the 16-s spectra (1a--d, 2a--d, and 3a--f) are given in
Table~\ref{tab:continua}.  The best-fitting parameters for the spectra
during dips 1 and 2 are very similar to the values obtained above for
the fits (with iron line included) of the longer-exposure pairs of
spectra, 1A/B and 2A/B respectively, from the same dips.  The
unabsorbed flux of the faint component relative to the unabsorbed flux
of bright component is 10--12\% in the fits of spectra 1a--d and 2a--d
and somewhat lower, 5.4\%, for spectra 3a--f.  The column density for
the faint component is again consistent with zero in all cases.
The Gaussian line parameters for these fits are listed in
Table~\ref{tab:gauss}, and indicate a line centered at
6.4--6.6~keV. The best-fitting line flux differs by a factor of two
among the three dips but is consistent, within the errors, with no
change.
For each set of spectra, the variable absorption column density of the
bright component is shown in Table~\ref{tab:absorp16s}. In each
case the column density increases from a relatively low value, $N_H =$
(4--9)$\times10^{22}$, at the start of a dip to very high values, $N_H >
10^{24}$, at the lowest part of the dips.

These fits show that the model developed above is quite successful in
reproducing the evolution of the spectrum through the dips.  This
justifies our assumption of common spectral values in the joint fits,
where only the absorption column density on the bright component
varies between all spectra of a given dip.

\section{Discussion}

During absorption dips in Cir~X-1, about 10\% of the total 2.5--25~keV
flux remains unabsorbed.  We conclude that the absorbing material does
not completely cover the X-ray source, and suggest that it is likely
that the material totally covers our line of sight to a compact X-ray
source during deep dips while a more extended source of X-rays remains
fully or nearly fully visible.
Similar behavior is observed during dips in other X-ray binaries.  For
example, an unabsorbed spectral component is detected during
absorption dips in the binary pulsar Her~X-1 and is attributed to
scattering in an extended corona (see \cite{stelzer99} and references
therein).  Likewise, curved CD/HID tracks similar to those we observe
in Cir~X-1 were observed in the black-hole candidates GRO~J1655-40 and
4U~1630-47 and were successfully modeled by absorption plus a faint
unabsorbed component (\cite{kuulkers98}; \cite{tomsick98}).

Our results are generally consistent with the \ASCA\ results of Brandt
et al.\ (1996\nocite{brandt96}, see above) for Cir~X-1, but with the
high-statistics \RXTE\ data, we are able to demonstrate the detailed
evolution of the continuum spectrum through selected high-to-low flux
transitions (dips).  The \ASCA\ spectra showed an iron K~edge due to
absorption with column densities near \tento{24}~cm$^{-2}$.  At
6.4~keV in the low-state \ASCA\ spectrum, an iron line was seen with a
flux of $\mscinot{(8.9^{+8.9}_{-5.6})}{-3}$
photons~cm$^{-2}$~s$^{-1}$, and in the high-state spectrum, an upper
limit of \scinot{4.5}{-3} photons~cm$^{-2}$~s$^{-1}$ was derived for
such a line.  This indicates that the equivalent width of the line is
much smaller in the high state, due to the brighter continuum, but
does not prove that the flux of the line was significantly smaller.
Our analysis shows that a prominent iron emission line is present
outside dips at a similar flux level, \scinot{(12-25)}{-3}
photons~cm$^{-2}$~s$^{-1}$ (see Table~\ref{tab:gauss}), as inside
dips.  The equivalent width is therefore much smaller outside dips,
as low as 44~eV, than during the deepest dips, where the equivalent
width is typically 410--940~eV\@.  Since scattering by an extended cloud
of material is probably responsible for the faint component, it is
likely that the iron K$\alpha$ emission is also produced in the
scattering medium (see below).

If the scattering medium was a spherically symmetric cloud around the
central X-ray source, scattering would not significantly alter the
total flux of the system but would simply enlarge the size of the
emission region for some fraction of the flux.
We observe that $\sim$10\% of the flux is in the diffuse component,
requiring a column density of almost \scinot{2}{23}~cm$^{-2}$ to
provide sufficient Thomson-scattering optical depth. 
The iron line associated with the scattered component indicates that
the scattering medium is not pure hydrogen.  Therefore, if the
scattering material is not highly ionized, the direct component should
show evidence for photoelectric absorption with a similar column
density.  The observed column density outside dips is about
\scinot{2}{22}~cm$^{-2}$---an order of magnitude less than required in
this simple model.  Thus we can conclude that the scattered component
is not produced in a spherical cloud of low-ionization state matter
which totally surrounds and covers the central source.

The column density derived from the strength of the faint component
can be used to estimate the iron K$\alpha$ fluorescent flux in the
scattering cloud.  Based on the spectrum we measure above the iron
K-shell ionization energy of 7.1~keV, a cloud of column density
\scinot{2}{23}~cm$^{-2}$ will photoelectrically absorb a flux of about
0.12 photons~cm$^{-2}$~s$^{-1}$.  The K$\alpha$ fluorescence yield for
iron is about 0.3 (\cite{bambynek72}), so we expect a flux in the
emission line of \scinot{36}{-3} photons~cm$^{-2}$~s$^{-1}$.  This is
actually somewhat higher, by a factor of 1.5--3, than the line fluxes
we have measured (Table~\ref{tab:gauss}); however, this simple model
illustrates that the iron K$\alpha$ emission in Cir~X-1 is naturally
explained by the strong scattered component revealed during dips.

We do not expect to find QPOs, all faster than 1~Hz in Cir~X-1, during
portions of dips when the direct component is almost entirely
absorbed, unless the scattering occurs in a region significantly
smaller than a light-second (roughly the size of the accretion disk,
$10^{10}$--$10^{11}$~cm, discussed below).
Power spectra during most of this 2-day observation show weak HBOs
near 30~Hz, broad NBOs at $\sim$4~Hz, or VLFN\@.  
No QPOs are detected during the lowest portions of absorption dips;
however, the upper limits on QPOs in these low count rate
circumstances do not allow us to exclude QPOs at a level similar to
that outside the dips.

The fact that absorption dips in Cir~X-1 mainly occur within a day of
phase zero and are then followed by significant flaring suggests that
the dips are associated with the mass transfer process.  Thus,
absorption might be due to the mass transfer stream itself, a bulge or
``splash point'' near the edge of an accretion disk due to collision
of an incoming stream with the disk, or a blob of outer disk material
protruding vertically into the line of sight.
Although absorption dips in Cir~X-1 show a wide range of profiles and
durations, for a brief deep dip we can attempt to estimate the size
and density of an absorber near the outer edge of an accretion
disk.

We first consider the simple case of blobs in Keplerian motion near
the outer edge of a disk.
Other than the 16.55~d orbital period of Cir~X-1, the binary
parameters, including a probable high eccentricity, are unknown.
However, the Keplerian velocity in the outer edge of a disk does not
depend strongly on the binary parameters, so we adopt values of
$M_{NS} \sim M_{donor} \sim 1.4$~\msun\ and eccentricity $\sim$ 0.8.
An outer disk radius of $R \sim 10^{10}$--$10^{11}$~cm, giving a
Keplerian velocity of order 1000~km~s$^{-1}$, would then be about
5--50\% of the radius of the instantaneous critical potential lobe of
the neutron star near periastron.
A typical brief deep dip such as Dip~1 lasts of order 100~s, implying
an azimuthal dimension of $\sim$\tento{10}~cm, or a significant
fraction of the accretion disk size.  If the radial and azimuthal
dimensions of the blob are similar, a column density of
\tento{24}~atoms~cm$^{-2}$ requires a number density in the blob of
$n\sim$\tento{14}~atoms~cm$^{-3}$.  
For typical luminosities of $L_x
\sim$\scinot{3}{38}~erg~s$^{-1}$~cm$^{-2}$ (determined from the
unabsorbed flux, assuming a distance to Cir X-1 of 8~kpc), the
ionization parameter $\xi = L_x / n R^2$ for a blob of this density
would be $\sim$\tento{3}--\tento{4}, which would in turn imply that
the blob of gas would be nearly fully ionized (\cite{kallman82}).
However, such a highly ionized absorber is inconsistent with the
strong Fe K edge near 7~keV observed during dips in Cir X-1.
The large size of the blobs in this picture relative to the disk size
might indicate that the blobs are actually azimuthally extended or
that dips, particularly the longer dips, are due to multiple blobs.
The Keplerian period of blobs in this picture is of order
\tento{3}~s; however, no recurrence of dips on this time scale is
apparent in the light curves, suggesting that such blobs may be
short-lived.

Alternately, we consider absorbers that are co-rotating with the
binary orbit, e.g, a stationary ``splash point'' on the outer edge of
the disk.  Using the same simple assumptions about the binary
parameters and disk size as above, we find an azimuthal velocity of
order 10~km~s$^{-1}$ for such an absorber.  This slower velocity would
require a much smaller absorber, with dimensions $\sim$\tento{8}~cm,
compared to a blob in a Keplerian orbit.  The corresponding number
density in the absorber would therefore be
$\sim$\tento{16}~atoms~cm$^{-3}$.  The ionization parameter would be
lower for these denser blobs, $\xi\sim$ 10--100, and the
absorbing gas would not be fully ionized, consistent with the
photoelectric absorption observed during dips (\cite{kallman82}).
In this picture, the fact that Cir~X-1 actually exhibits many dips
during a couple days of each binary cycle suggests a stationary (in
the rotating frame) structure in the accretion disk (i.e., a
stationary wave) or else complex structure (e.g., shocks or
turbulence) in the accretion stream.

\begin{figure*}
\begin{centering}
\PSbox{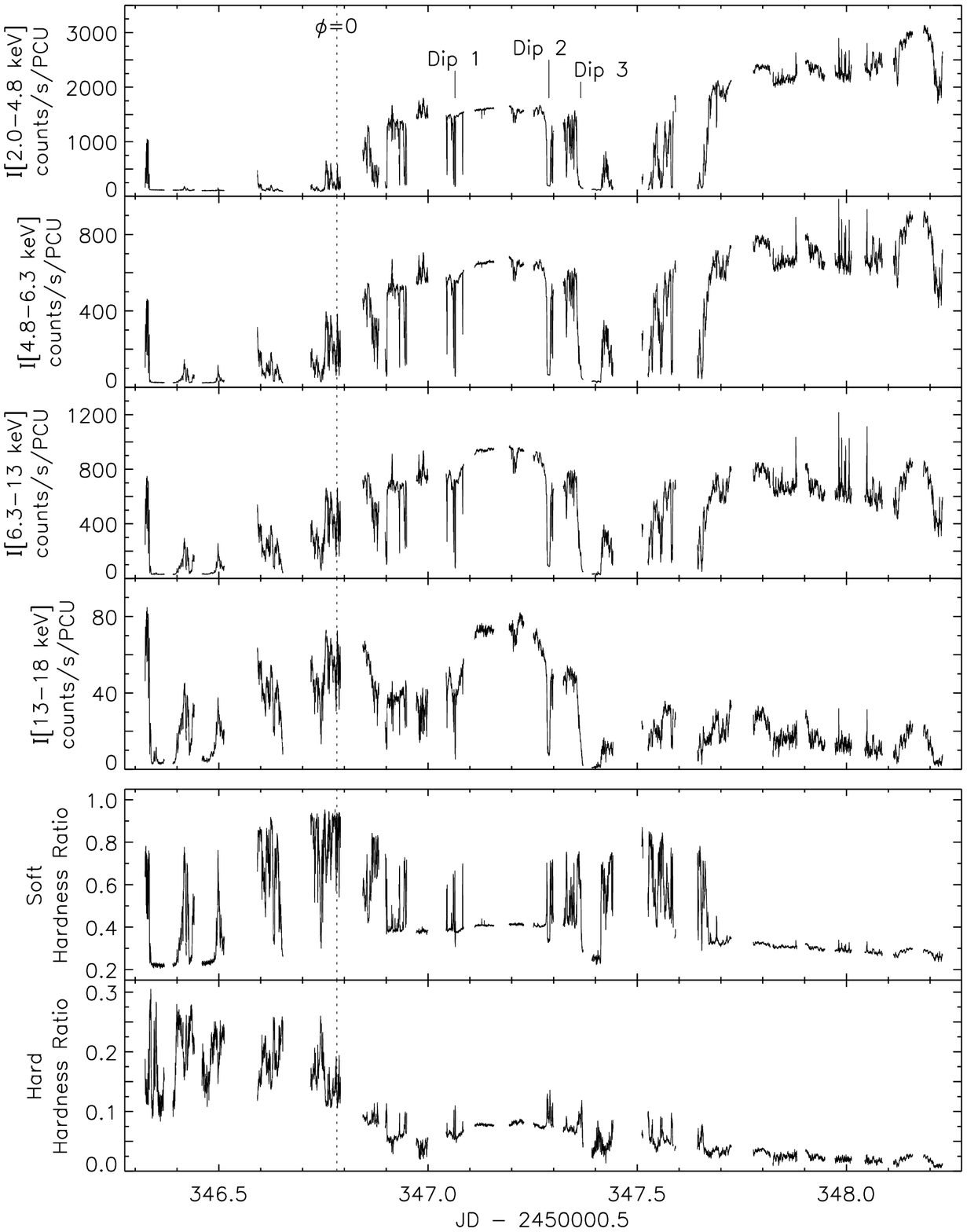 hscale=90 vscale=90 hoffset=-10}{6.1in}{7.7in}
\caption{
Light curves in four energy channels and two hardness ratios for PCA
observations of Cir~X-1 from 1996 September~20--22, covering a two-day
period around phase zero ($\phi=0$). Three dips have been identified
for further study. Each point represents 16~s of background-subtracted
data from all five PCA detectors. Ratios of the intensities produce soft
(\hardness{2.0}{4.8}{4.8}{6.3}) and hard (\hardness{6.3}{13}{13}{18})
hardness ratios. The intensity levels of the segment at
day~$\sim$347.2 (after Dip~1) are close to the level in each band
during the quiescent phases of the orbit. 
}
\label{fig:48hr}
\end{centering}
\end{figure*}

\begin{figure*}
\begin{centering}
\PSbox{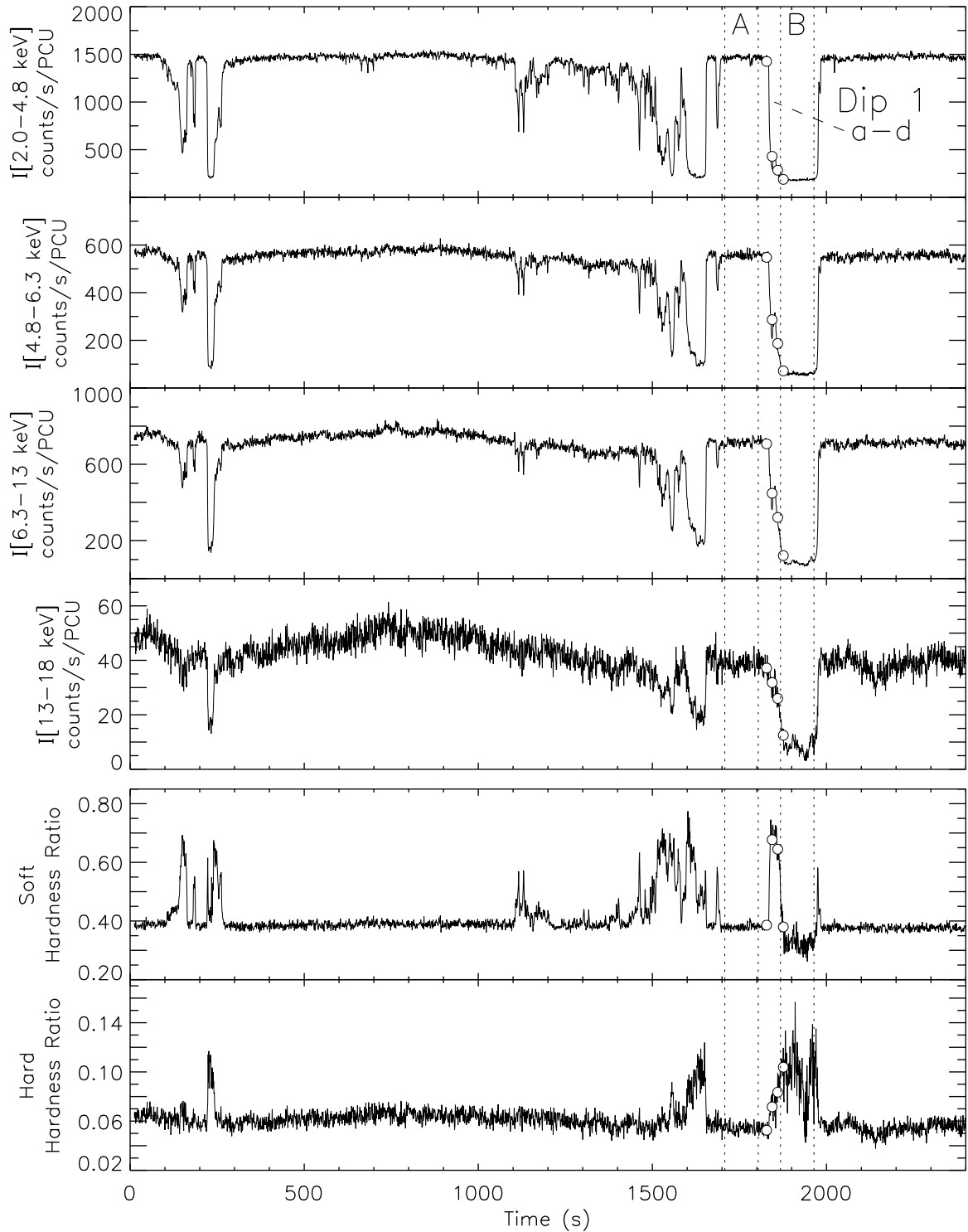 hscale=90 vscale=90 hoffset=-10}{6.1in}{7.7in}
\caption{
Expanded 2400-s segment (1-s bins) of the light curves and hardness
ratios in Fig.~\protect{\ref{fig:48hr}} including Dip~1.  Time
zero corresponds to day~347.043 in
Fig.~\protect{\ref{fig:48hr}}.  Energy spectra were extracted
from two 96-s time segments (A and B), indicated by dotted vertical
lines, and four 16-s segments (a--d) indicated by circles.
\label{fig:dip1lc}
}
\end{centering}
\end{figure*}

\begin{figure*}
\begin{centering}
\PSbox{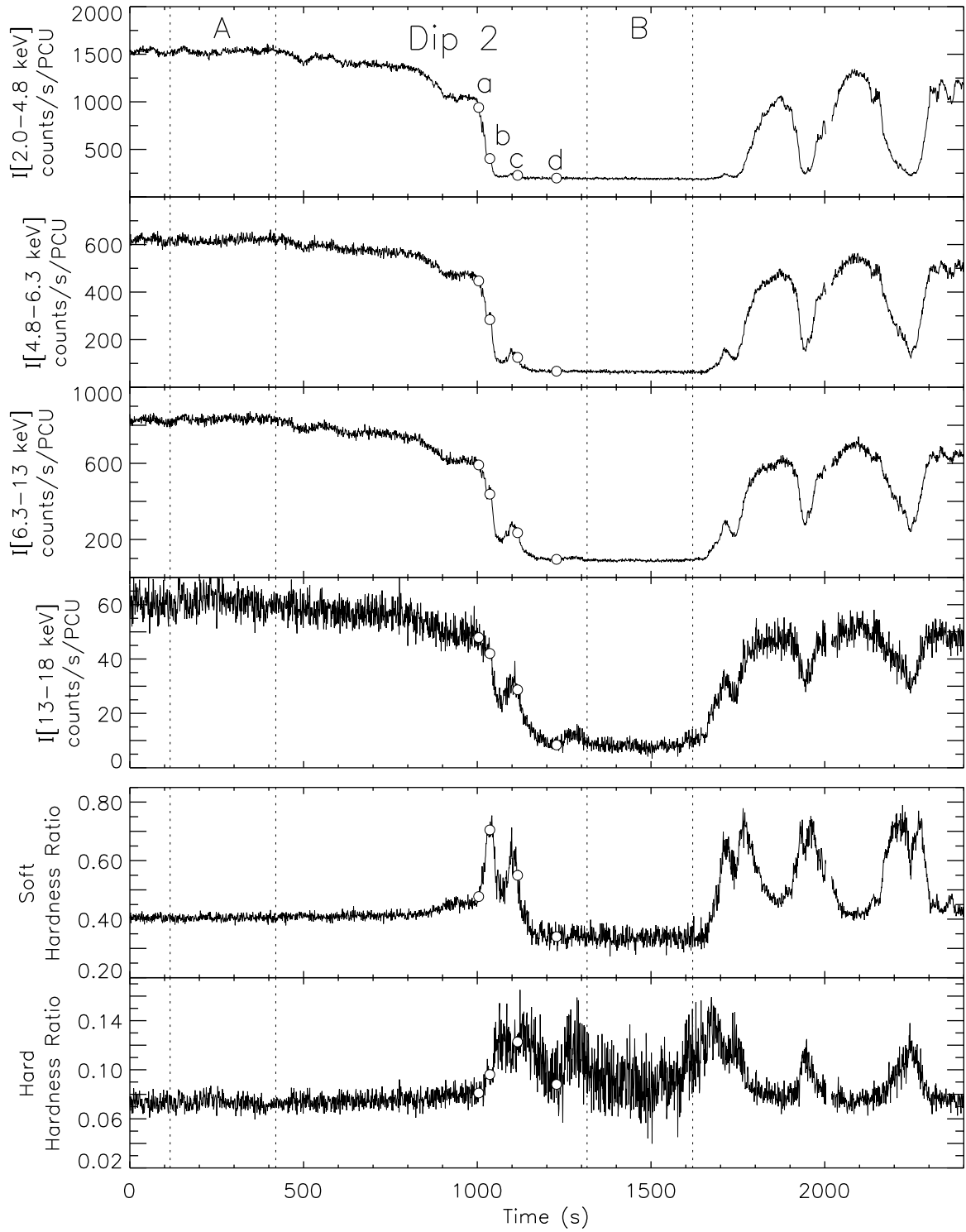 hscale=90 vscale=90 hoffset=-10}{6.1in}{7.7in}
\caption{
Expanded 2400-s segment (1-s bins) of the light curves and hardness
ratios in Fig.~\protect{\ref{fig:48hr}} including Dip~2. Time
zero corresponds to day~347.272 in
Fig.~\protect{\ref{fig:48hr}}.  Energy spectra were extracted
from two 304-s time segments (A and B) and four 16-s segments (a--d).
\label{fig:dip2lc}
}
\end{centering}
\end{figure*}

\begin{figure*}
\begin{centering}
\PSbox{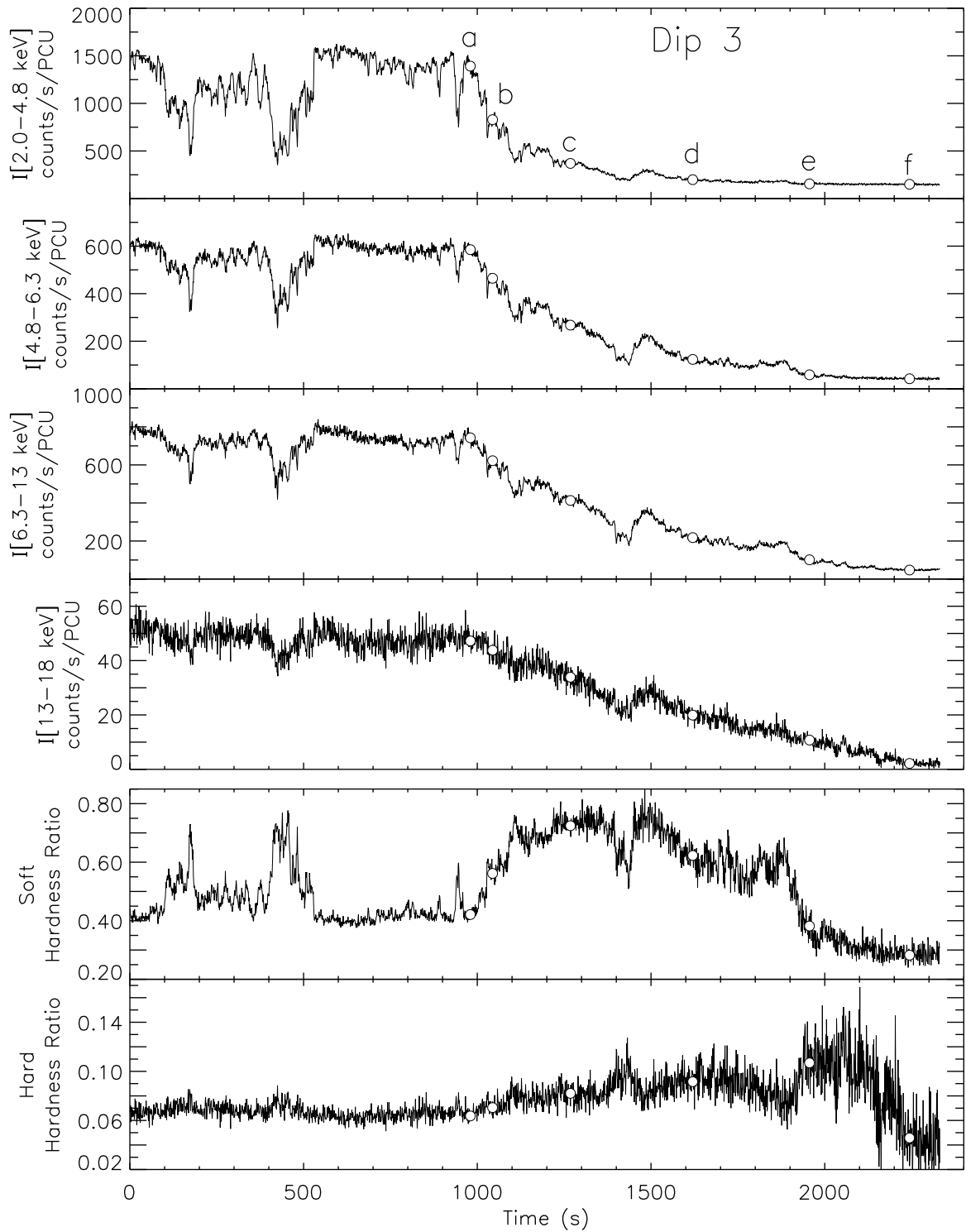 hscale=90 vscale=90 hoffset=-10}{6.1in}{7.7in}
\caption{
Expanded 2400-s segment (1-s bins) of the light curves and hardness
ratios in Fig.~\protect{\ref{fig:48hr}} including Dip~3. Time
zero corresponds to day~347.344 in
Fig.~\protect{\ref{fig:48hr}}.  Energy spectra were extracted
from six 16-s time segments (a--f).
\label{fig:dip3lc}
}
\end{centering}
\end{figure*}

\begin{figure*}
\begin{centering}
\PSbox{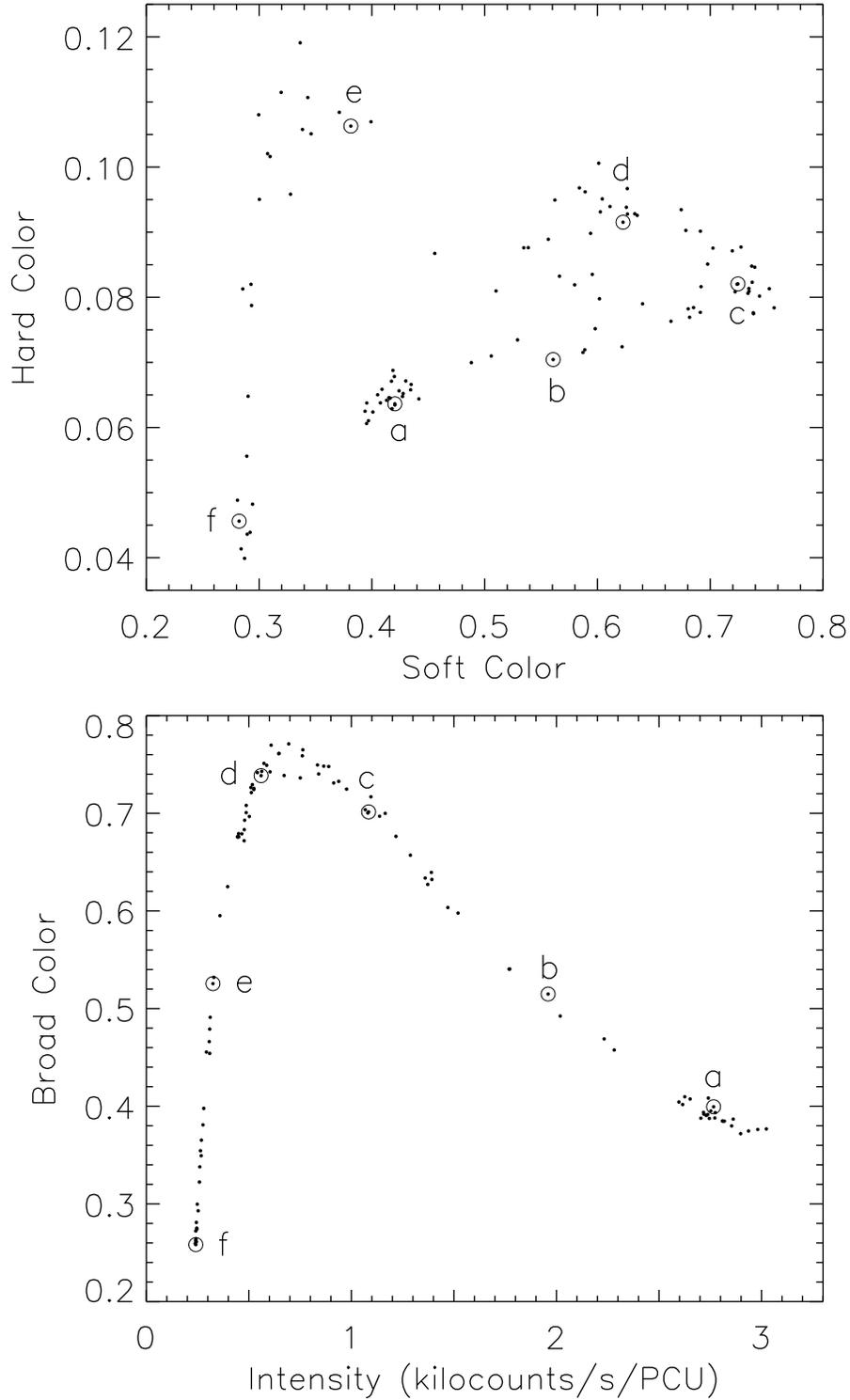 hscale=90 vscale=90 voffset=-16 hoffset=-35}{4.5in}{7.5in}
\caption{
Color-color and hardness-intensity diagrams for time 600~s through
2400~s of Fig.~\protect{\ref{fig:dip3lc}}, during which the
intensity gradually transitioned from the non-dip baseline to the
bottom of a dip. Circled points labeled `a--f' correspond to the six
16-s time segments identified in
Fig.~\protect{\ref{fig:dip3lc}} and used to construct energy
spectra.  Intensity is \intens{2.0}{18}, and the hardness ratios are
defined as soft color: \hardness{2.0}{4.8}{4.8}{6.3}, hard color:
\hardness{6.3}{13}{13}{18}, and broad color:
\hardness{2.0}{6.3}{6.3}{18}.  Each point represents 16~s of
background-subtracted data from all five PCA detectors.
\label{fig:cc_hid_85911900_2ks}
}
\end{centering}
\end{figure*}

\begin{figure*}
\begin{centering}
\PSbox{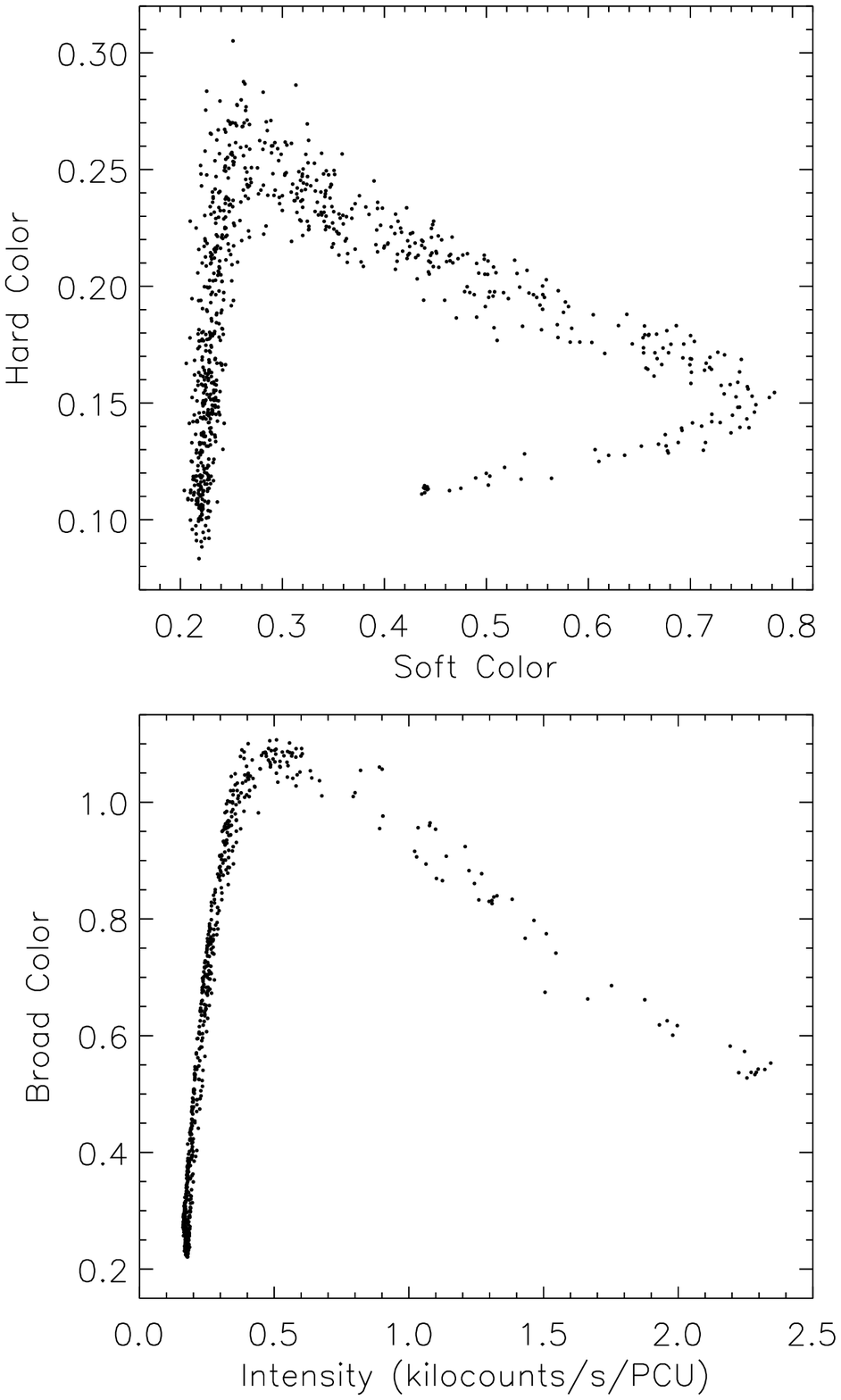 hscale=90 vscale=90 voffset=-16 hoffset=-35}{4.5in}{7.5in}
\caption{
Color-color and hardness-intensity diagrams for the first three segments in
Fig.~\protect{\ref{fig:48hr}} (day 346.31--346.52), during
which Cir~X-1 was in an extended low/dip state. The data were
constructed as in Fig.~\protect{\ref{fig:cc_hid_85911900_2ks}}.
\label{fig:cc_hid_85823000_17ks}
}
\end{centering}
\end{figure*}

\begin{figure*}
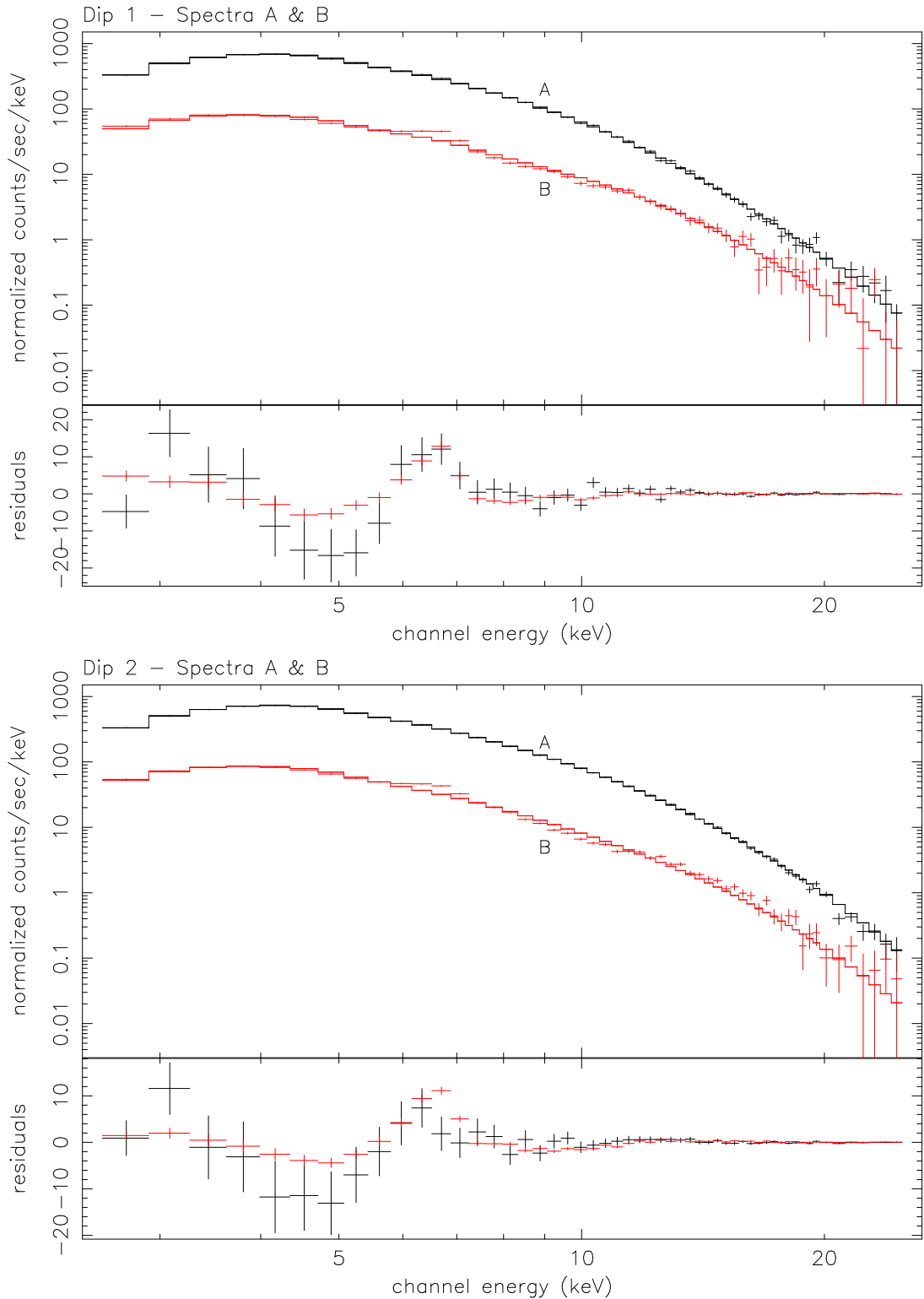

\begin{centering}
\PSbox{fig7a.eps hscale=62 vscale=62 angle=270 voffset=350 hoffset=-25}{5.6in}{3.9in}
\PSbox{fig7b.eps hscale=62 vscale=62 angle=270 voffset=340 hoffset=-25}{5.6in}{3.9in}
\caption{ 
Top: Spectral fits for 96-s segments prior to Dip~1 (spectrum~A) and
during Dip~1 (spectrum~B).  Bottom: Spectral fits for 304-s segments
prior to Dip~2 (spectrum~A) and during Dip~2 (spectrum~B).
The data shown are from PCU~0 only. 
\label{fig:dips1and2_nogauss}
}
\end{centering}
\end{figure*}

\begin{figure*}
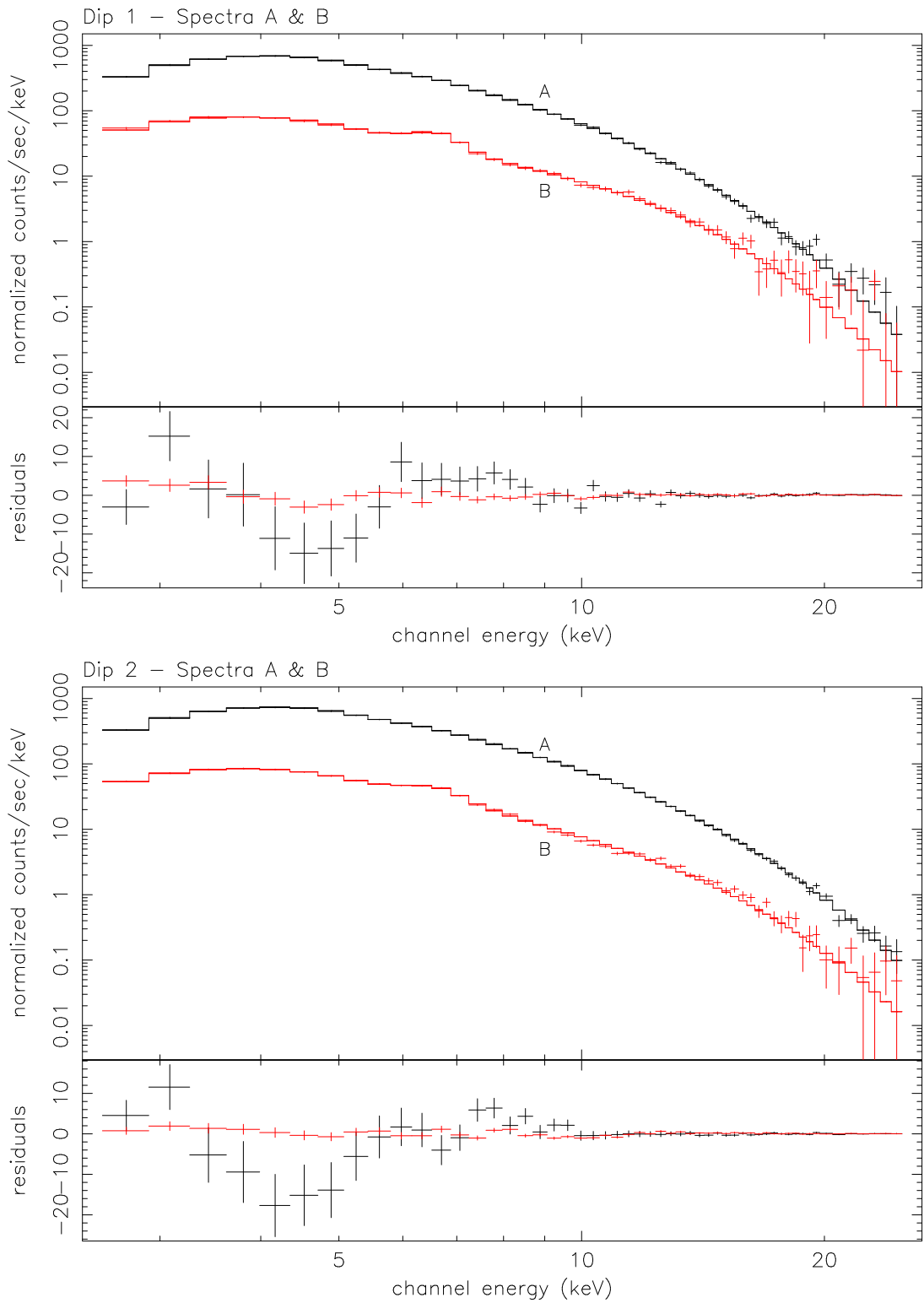

\begin{centering}
\PSbox{fig8a.eps hscale=62 vscale=62 angle=270 voffset=350 hoffset=-25}{5.6in}{3.9in}
\PSbox{fig8b.eps hscale=62 vscale=62 angle=270 voffset=340 hoffset=-25}{5.6in}{3.9in}
\caption{ 
Same spectra and model as in Fig.~\protect{\ref{fig:dips1and2_nogauss}}, 
but with a Gaussian emission-line component included to fit the peaked 
residuals near 6.5~keV.
\label{fig:dips1and2_gauss}
}
\end{centering}
\end{figure*}

\begin{figure*}
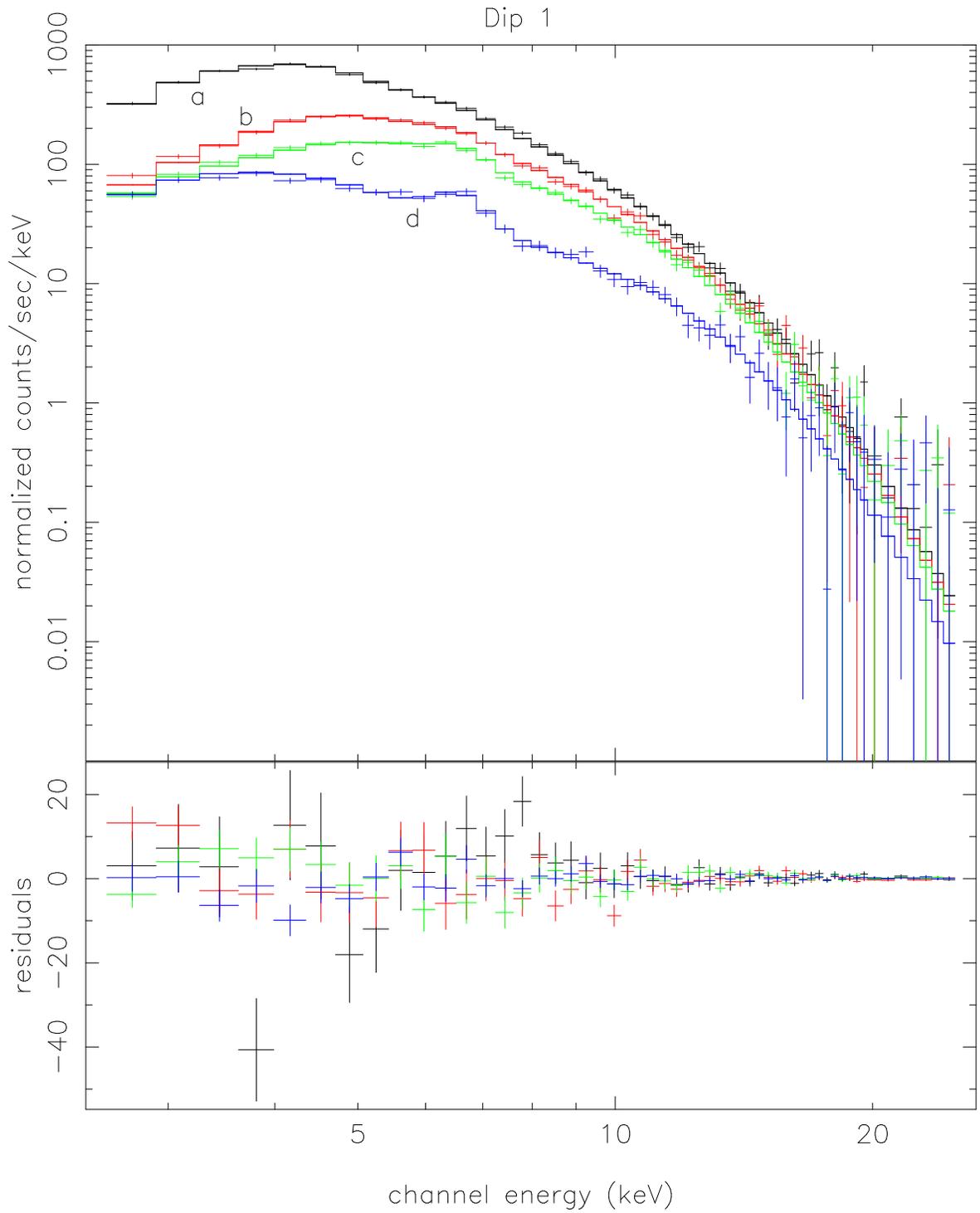

\begin{centering}
\PSbox{fig9.eps hscale=90 vscale=90 voffset=-18 hoffset=-40}{6.1in}{7.4in}
\caption{
Spectral fits for four 16-s segments (a--d) during the decline into
Dip~1. Only the column density of the bright component varies between
the four jointly-fit curves. The data shown are from PCU~0 only.
\label{fig:dip1_four16s}
}
\end{centering}
\end{figure*}

The chaotic nature of the PCA light curves during dips
(Figs.~\ref{fig:48hr}--\ref{fig:dip3lc}) indicate that we are
observing fine structure (temporal or geometrical) of the obscuring
material.  The true nature of the absorbing material is likely to be
as complex as the light curves themselves, reflecting the turbulent
conditions associated with mass transfer in an eccentric binary.  In
Cir~X-1 the accretion disk probably experiences a high degree of
disturbance due to a high eccentricity with enhanced mass transfer
occurring near periastron.

\acknowledgements

We acknowledge the \RXTE\ teams at MIT \& GSFC for
their support.  In particular we thank E.~Morgan, R.~Remillard,
W.~Cui, and D.~Chakrabarty for useful discussions.  Support for this
work was provided through NASA Contract NAS5-30612.


\newpage

\begin{figure*}
\begin{centering}
\PSbox{fig10.eps hscale=90 vscale=90 voffset=-18 hoffset=-40}{6.1in}{7.4in}
\caption{
Spectral fits for four 16-s segments (a--d) during the decline into
Dip~2.  See the caption of Fig.~\protect{\ref{fig:dip1_four16s}}.
\label{fig:dip2_four16s}
}
\end{centering}
\end{figure*}

\begin{figure*}
\begin{centering}
\PSbox{fig11.eps hscale=90 vscale=90 voffset=-18 hoffset=-40}{6.1in}{7.4in}
\caption{
Spectral fits for six 16-s segments (a--d) during the decline into
Dip~3.  See the caption of Fig.~\protect{\ref{fig:dip1_four16s}}.
}
\label{fig:dip3_six16s}
\end{centering}
\end{figure*}

\newpage

\begin{deluxetable}{ccccccccc}
\renewcommand{\arraystretch}{1.5}
\tablewidth{0pt}
\footnotesize
\tablecaption{ \label{tab:continua}
Joint fit parameters of the bright and faint spectral components
during dips. }
\tablehead{ 
\colhead{} & 
\colhead{} & 
\multicolumn{2}{c}{Disk Blackbody} &
\multicolumn{2}{c}{Blackbody} &
\multicolumn{2}{c}{Faint Component}  & 
\colhead{}
\nl 
\colhead{ Joint } & 
\colhead{ Iron } &
\colhead{ $kT_{in}$ } & 
\colhead{ $R_{in}\cos^{1/2}\theta$\tablenotemark{c} } & 
\colhead{ $kT$ } & 
\colhead{ $R$\tablenotemark{d} } & 
\colhead{} & 
\colhead{ $N_H^{ (2) }/10^{22}$~\tablenotemark{f} } & 
\colhead{} 
\nl
\colhead{ fit\tablenotemark{a} } & \colhead{ Line\tablenotemark{b} } &
\colhead{ (keV)} & \colhead{ (km) } &
\colhead{ (keV)} & \colhead{ (km) } & 
\colhead{ $f$~\tablenotemark{e} } & \colhead{(cm$^{-2}$) } & 
\colhead{ $\chi^2_r$~\tablenotemark{g} } 
}
\startdata
1A--B & no &
	$1.51^{+0.10}_{-0.17}$ & $16.94^{+3.96}_{-2.20}$ &
	$2.20^{+0.62}_{-0.42}$ & $3.77^{+3.33}_{-2.55}$ &
	$0.107^{+0.005}_{-0.006}$ &
	$0.00^{+0.07}_{-0.00}$ & 
	3.89--4.07 \nl
2A--B & no &
	$1.51^{+0.04}_{-0.06}$ & $17.54^{+1.25}_{-0.87}$ & 
	$2.32^{+0.16}_{-0.15}$ & $3.47^{+0.86}_{-0.76}$ & 
	$0.103^{+0.003}_{-0.003}$ &
	$0.03^{+0.28}_{-0.03}$ & 
	7.58--7.84 \nl
\hline
1A--B & yes & 
	$1.23^{+0.22}_{-0.15}$ & $25.32^{+8.56}_{-7.03}$ & 
	$1.79^{+0.33}_{-0.16}$ & $7.87^{+3.34}_{-4.14}$ & 
	$0.096^{+0.007}_{-0.006}$ &
	$0.00^{+0.10}_{-0.00}$ & 
	1.44--1.62 \nl
2A--B & yes &
	$1.35^{+0.06}_{-0.05}$ & $22.45^{+1.83}_{-1.69}$ & 
	$2.10^{+0.12}_{-0.09}$ & $5.23^{+0.92}_{-0.93}$ & 
	$0.092^{+0.002}_{-0.002}$ &
	$0.00^{+0.10}_{-0.00}$ & 
	1.87--2.15 \nl
\hline
1a--d & yes &
	$1.18^{+0.19}_{-0.16}$ & $27.04^{+9.39}_{-6.33}$ & 
	$1.72^{+0.22}_{-0.12}$ & $8.22^{+2.69}_{-3.03}$ & 
	$0.103^{+0.009}_{-0.011}$ &
	$0.00^{+0.44}_{-0.00}$ &
	0.99--1.20 \nl
2a--d & yes & 
	$1.26^{+0.14}_{-0.14}$ & $25.12^{+7.66}_{-5.20}$ & 
	$2.08^{+0.25}_{-0.15}$ & $5.46^{+1.08}_{-1.08}$ & 
	$0.117^{+0.013}_{-0.014}$ &
	$0.15^{+0.70}_{-0.15}$ & 
	0.80--1.12 \nl
3a--f & yes & 
	$1.07^{+0.08}_{-0.09}$ & $42.66^{+8.34}_{-6.60}$ & 
	$1.74^{+0.09}_{-0.09}$ & $9.23^{+1.41}_{-1.32}$ & 
	$0.054^{+0.005}_{-0.005}$ &
	$0.00^{+0.14}_{-0.00}$ & 
	1.27--1.34 \nl
\enddata
\tablenotetext{a}{
Joint fits of spectra from outside (A) and inside (B) dips 1 and~2
and for the four to six 16-s spectra (a--f) during dips 1, 2, and~3
(using the model given in equation~\protect{\ref{eq:dip_model}}). Only
the column density ($N_H^{(1)}$) of the bright component is allowed to
vary between spectra (see Tables~\protect{\ref{tab:absorpAB}}
and~\protect{\ref{tab:absorp16s}}). Errors quoted are 90\%
confidence limits for a single parameter ($\Delta\chi^2=2.7$).}
\tablenotetext{b}{
Gaussian emission line at 6.4--6.6~keV was included in the final
five fits (see Table~\protect{\ref{tab:gauss}}). }
\tablenotetext{c}{
Inner radius of the accretion disk (times
$\cos^{1/2}\theta$, where $\theta$ is the angle between the normal to
the disk and the line of sight) for a distance of 8~kpc.}
\tablenotetext{d}{Blackbody radius for a distance of 8~kpc.}
\tablenotetext{e}{
Ratio of the unabsorbed flux of the faint component to the unabsorbed 
flux of the bright component.
}
\tablenotetext{f}{
Absorption column density of the faint component. See 
Table~\protect{\ref{tab:absorpAB}} for the variable absorption of 
the bright component.}
\tablenotetext{g}{
$\chi^2/dof$, where $dof =$ the number of spectral bins (52--54 per spectrum) 
minus the number of fit parameters. The range of values represents the fits 
for PCUs 0, 1, \& 4.}
\renewcommand{\arraystretch}{1}
\end{deluxetable}

\begin{deluxetable}{cccc}
\renewcommand{\arraystretch}{1.5}
\tablewidth{0pt}
\tablecaption{ \label{tab:absorpAB}
Absorption column density $N_H^{(1)}$ of the bright component outside
(A) and inside (B) Dips 1 and~2.  }
\tablehead{
\colhead{} & \colhead{ $N_H^{(1)}/10^{22}$ } &
\colhead{} & \colhead{ $N_H^{(1)}/10^{22}$ } \nl
\colhead{Spectrum} & \colhead{(cm$^{-2}$)} & 
\colhead{Spectrum} & \colhead{(cm$^{-2}$)} 
}
\startdata
\multicolumn{4}{c}{No Emission Line} \nl
1A & $2.83^{+0.57}_{-0.43}$ & 2A & $3.50^{+0.29}_{-0.34}$ 	  \nl
1B & $176^{+12}_{-13}$ & 2B & $306^{+21}_{-19}$ \nl
\cline{1-4}
\multicolumn{4}{c}{With Emission Line\tablenotemark{a}} \nl
1A & $3.81^{+0.71}_{-0.81}$ & 2A & $4.30^{+0.26}_{-0.26}$ \nl
1B & $184^{+15}_{-14}$ & 2B & $283^{+16}_{-14}$ \nl
\enddata
\tablenotetext{a}{
A Gaussian emission line at 6.4--6.6~keV was included, with the
same parameters outside and inside the dips (see
Table~\protect{\ref{tab:gauss}}). }
\renewcommand{\arraystretch}{1}
\end{deluxetable}

\begin{deluxetable}{cccc}
\renewcommand{\arraystretch}{1.5}
\tablewidth{0pt}
\tablecaption{ \label{tab:gauss}
Gaussian emission line parameters for the joint fits of dip spectra.
}
\tablehead{
\colhead{ Joint } & \colhead{ $E$ }  & \colhead{ $\sigma$\tablenotemark{a} } & \colhead{ Line flux} \nl
\colhead{ Fit }   & \colhead{ (keV) } & \colhead{ (keV) }   & \colhead{ (photons~cm$^{-2}$~s$^{-1}$)}
}
\startdata
1A--B & $6.59^{+0.12}_{-0.13}$ & $0.18^{+0.28}_{-0.18}$ & $0.018^{+0.007}_{-0.004}$ \nl
2A--B & $6.58^{+0.06}_{-0.07}$ & $0.29^{+0.10}_{-0.13}$ & $0.016^{+0.002}_{-0.002}$ \nl
\hline
1a--d & $6.59^{+0.16}_{-0.15}$ & $0.13^{+0.29}_{-0.13}$ & $0.015^{+0.002}_{-0.002}$ \nl
2a--d & $6.45^{+0.16}_{-0.20}$ & $0.47^{+0.29}_{-0.32}$ & $0.025^{+0.012}_{-0.008}$ \nl
3a--f & $6.46^{+0.23}_{-0.27}$ & $0.39^{+0.32}_{-0.39}$ & $0.012^{+0.004}_{-0.003}$ \nl
\enddata
\tablenotetext{a}{
Gaussian width.}
\renewcommand{\arraystretch}{1}
\end{deluxetable}

\begin{deluxetable}{cccccc}
\renewcommand{\arraystretch}{1.5}
\tablewidth{0pt}
\tablecaption{ \label{tab:absorp16s} 
Effective hydrogen column density responsible for the variable
absorption of the bright component during 16-s spectra from dips 1, 2,
and~3.  }
\tablehead{
\colhead{} & \colhead{$N_H^{(1)}/10^{22}$} &
\colhead{} & \colhead{$N_H^{(1)}/10^{22}$} & 
\colhead{} & \colhead{$N_H^{(1)}/10^{22}$} \nl
\colhead{Spectrum} & \colhead{(cm$^{-2}$)} & 
\colhead{Spectrum} & \colhead{(cm$^{-2}$)} & 
\colhead{Spectrum} & \colhead{(cm$^{-2}$)} 
}
\startdata
1a & $4.16^{+1.10}_{-0.75}$ & 2a & $8.84^{+1.37}_{-1.09}$ & 3a & $8.31^{+1.09}_{-0.96}$ \nl
1b & $25.8^{+1.6}_{-1.3}$ & 2b & $25.8^{+1.4}_{-1.5}$ & 3b & $15.7^{+1.0}_{-1.1}$ \nl
1c & $43.6^{+1.7}_{-2.1}$ & 2c & $64.0^{+2.4}_{-2.5}$ & 3c & $33.2^{+1.4}_{-1.5}$\nl
1d & $129^{+8}_{-8}$ & 2d &$242^{+83}_{-52}$  & 3d & $66.7^{+1.6}_{-1.5}$\nl
   &  &    &  & 3e & $127^{+8}_{-7}$ \nl
   &  &    &  & 3f & $>\mscinot{2}{4}$ \nl
\enddata
\renewcommand{\arraystretch}{1}
\end{deluxetable}

\end{document}